# A Novel Matrix Representation of Discrete Biomedical Signals

Aditya Ramesh,
Electrical and Electronics Engineering, National Institute of Technology, Karnataka, Surathkal,
Mangaluru 575025, India. (e-mail: aditya_2806@ieee.org)

Anagh Pathak, Kaushik Majumdar
Systems Science and Informatics Unit, Indian Statistical Institute, 8th Mile, Mysore Road, Bangalore
560059, India (e-mails: pathak02@gmail.com, kmajumdar@isibang.ac.in)

*Abstract*— **In this work we propose a novel symmetric square matrix representation of one or more digital signals of finite equal length. For appropriate window length and sliding paradigm this matrix contains useful information about the signals in a two dimensional image form. Then this representation can be treated either as an algebraic matrix or as a geometric image. We have shown applications of both on human multichannel intracranial electroencephalogram (iEEG). In the first application we have shown that for certain patients the highest eigenvalue of the matrix obtained from the epileptic focal channels goes up during a seizure. The focus of this paper is on an application of the second concept, by which we have come up with an automatic seizure detection algorithm on a publicly available benchmark data. Except for delay in detection in all other aspects the new algorithm outperformed the detection performance based on a support vector machine based algorithm. We have also indicated how this sparse random matrix representation of brain electrical signals can encode the activities of the brain.**

## I. Introduction

MATRIX representation of discrete biomedical signals is not uncommon. Often, multichannel electroencephalogram (EEG) signal is represented in matrix form [1], [2]. Once represented in matrix form various multivariate analyses like Independent Component Analysis (ICA) [3], Principal Component Analysis (PCA) [4], singular value decomposition (SVD) of EEG signals [5], EEG source localization by Moore-Penrose pseudo inverse [6], etc. Cortical source localization problem of scalp EEG has two components, namely forward problem (see ref. [7] for a review) and inverse problem (see ref. [8] for a review). Matrix representation of the EEG signals is essential in both the problems. In many instances multichannel EEG is studied simultaneously with electrocardiogram (ECG) and electrooculogram (EOG) signals, from which artifacts due to the two are to be identified and eliminated. This is typically done by linear decomposition techniques such as, ICA, PCA, SVD etc. The first step of such a process is to represent EEG, ECG and EOG signals within a single matrix.

Biomedical signals carry signatures of physiological events. Detection of events is therefore an important step in biomedical signal analysis [9]. Autocorrelation of a biomedical signal has been used as an event detector [10] and so also cross correlation [11]. The part of a signal related to a specific event is often referred to as an *epoch* [9]. Autocorrelation becomes high during an epoch and cross correlation becomes high between the two signals when they both contain the same epoch. This is the way epochs are identified in signals by the two methods. In this work we will show that similar effects are possible by multiplication of a (one-channel or multichannel) signal matrix with its transpose. We will show how events like epileptic seizures can be detected in the human depth EEG signals by this method.

In the central nervous systems (CNSs) different tasks are performed in different specialized areas. One particular event therefore can have a modular effect on the CNS of an animal, that is, different specialized parts of the CNS can be activated by the event and simultaneity of those activations will represent the event in the CNS. In order to monitor the CNS in response to that event multiple signals from different parts of the CNS may have to be studied for simultaneity, similarity, correlation, etc. Different quantitative measures have been designed for numerically measuring these attributes, such as, phase synchronization [12], [13] amplitude correlation [14], peak synchronization [15], etc. In this work we will show that multichannel signal matrix multiplication with its transpose is able to capture near simultaneous epochs in those channels, which are important signatures of healthy or pathological functioning of the CNS.

In the next section we will describe a new matrix representation of one or more time domain digital signals as if the matrix will work as a two dimensional image containing information from those signals. In section III we will present an application of this representation in automatic detection of epileptic seizures in iEEG signals recorded from patients with epilepsy. The last section will contain concluding remarks and indications to a few possible future directions.

## II. MATRIX REPRESENTATION

Let single or multichannel digitized signal be represented in matrix form in the usual way. Let the columns be the channels and the rows be the samples. Let $\mathbf{A}$ be a $T \times N$ signal matrix, where $T$ is the number of time points and $N$ is the number of one dimensional time domain digital signals. Then the $T \times T$ square matrix $\mathbf{M}$ is given by

$$\mathbf{M} = \mathbf{A}\mathbf{A}^T, \qquad (1)$$

where $\mathbf{A}^T$ denotes the transpose of $\mathbf{A}$. $\mathbf{M}$ is the new matrix representation of the single channel or multichannel signal. Note that dimension of $\mathbf{M}$ does not depend on $N$. Now the question is what does $\mathbf{M}$ mean? We will explain it with an example.

Let $\mathbf{A}$ be a $3 \times 2$ matrix as following

$$\mathbf{A} = \begin{bmatrix} s_1(t_1) & s_2(t_1) \\ s_1(t_2) & s_2(t_2) \\ s_1(t_3) & s_2(t_3) \end{bmatrix}, \qquad (2)$$

where $s_1()$ and $s_2()$ are two discrete signals and $t_1$, $t_2$ and $t_3$ are three successive time points. Then

$$\mathbf{M} = \begin{bmatrix} \{s_1(t_1)\}^2 + \{s_2(t_1)\}^2 & s_1(t_1)s_1(t_2) + s_2(t_1)s_2(t_2) \\ s_1(t_1)s_1(t_2) + s_2(t_1)s_2(t_2) & \{s_1(t_2)\}^2 + \{s_2(t_2)\}^2 \\ s_1(t_1)s_1(t_3) + s_2(t_1)s_2(t_3) & s_1(t_2)s_1(t_3) + s_2(t_2) + s_2(t_3) \end{bmatrix}$$

Clearly, each element in $\mathbf{M}$ is a summation of product of sample values from different one dimensional signals. Except in the diagonal of $\mathbf{M}$, in all other elements the product of samples are at different time points in the same signal. How far apart these time points will be that depends on the value of $T$. In the diagonal it is sum of squares of samples across different signals but all at the same time point.

How this notion can be useful has been shown in Fig 1 and Fig 2. Fig 1 is an image of $\mathbf{M}$, where $\mathbf{A}$ is a $1000 \times 3$ matrix. $1000$ successive samples have been taken from each of the $3$ focal EEG signals during the progression of an epileptic seizure. During seizure the signal amplitude goes up in the focal EEG (that is, the EEG signal collected from where the seizure is being generated). Statistically most of the sample values across all the signals remain high during that time. So their sum of product also remains high. This is clearly visible in Fig 1. On the other hand Fig 2 has been generated by the same three signals of same duration, but during an interictal period, that is, the period in between two successive seizures. During this time much lesser number of sample values reach as high as during the seizure, which is clearly reflected in Fig 2.

The result can be further improved by filtering the signals, by making the sample values all nonnegative (by adding the estimated global minimum of the signal to the sample values), by applying temporal difference operators, which are known to enhance the contrast between seizure and background signals [16], [17], etc. In biomedical signals, such as, EEG, MEG, fMRI, in which simultaneous high amplitude among the signals collected from different regions may indicate important events, representation of signals of interest in the form of equation (1) will be useful for further processing.

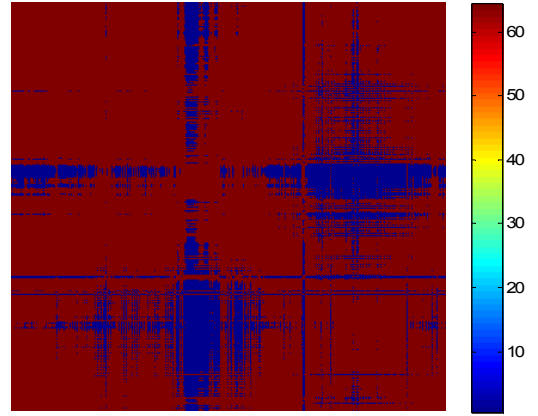

Fig 1. Topographic image generated by 1000 time points of three focal EEG signals (sample frequency 256 Hz) during the progression of a seizure.

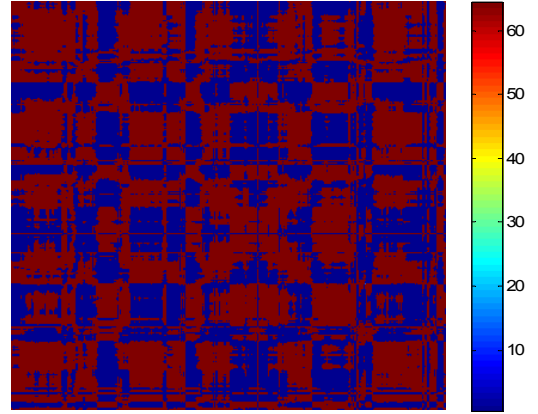

Fig 2. Topographic image generated by 1000 time points of the same three focal EEG signals as in Fig 1 during interictal period.

One advantage of equation (1) is getting a two dimensional image out of one or more one dimensional signals. Standard image processing techniques, such as edge detection, can be applied on the image of $\mathbf{M}$ to decipher information about epochs in signals in $\mathbf{A}$.

$\mathbf{M}$ can also be seen from another angle. Each of the digitized EEG signals $s_i(t)$ may be treated as a collection of random variables (for a signal segment of 1000 samples there are 1000 random variables), that is, as a stochastic process. Then $\mathbf{M}$ is a random matrix [18]. For some patients the highest eigenvalue of $\mathbf{M}$ has higher values during the seizures than in the interictal periods (Fig 3). This is likely to happen when the electrographic seizure is manifested strongly and simultaneously in all the focal channels. $\mathbf{M}$ may contain important information about the dynamics of seizure, but here we will only explore the image of $\mathbf{M}$.

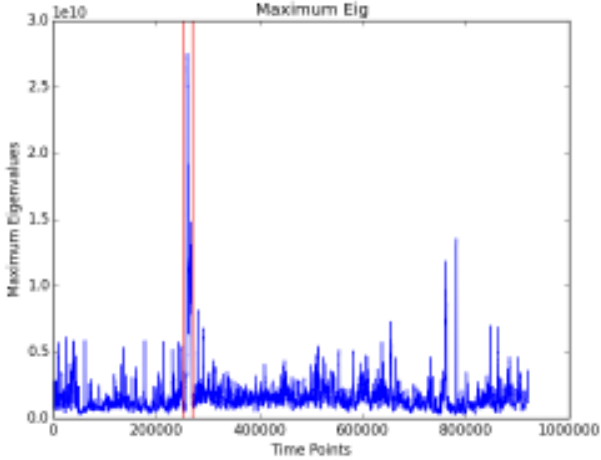

Fig 3. The plot of maximum eigenvalue of the matrix $\mathbf{M}$ as described above, obtained from one hour 3-focal channel recording of iEEG signals before, during and after an epileptic seizure. Start and end of seizure have been marked by vertical lines. Here 256 time points = 1 second.

### III. APPLICATION

Here we will present one application of the representation (1) in automatic seizure detection in intracranial EEG (iEEG) signals emanating out of the seizure foci or the seizure onset zone (SOZ).

#### A. Data

We have used iEEG signals collected from 21 patients with epilepsy in the University Hospital of Freiburg, Freiburg, Germany [18]. This data was made available freely to the researchers from any part of the world by the Freiburg Seizure Prediction Project of the Albert-Ludwig-Universitat Freiburg, Germany {18] (we downloaded it in 2009) and is particularly suitable for benchmarking various algorithms for seizure detection [16], [17] and prediction [20], [21].

The iEEG data collected using Neurofile NT digital video EEG system (It-med, Usingen, Germany) with 128 channels, at 256 Hz sampling rate. A 16 bit analog to digital converter was used for sample value encoding. We were only given iEEG from six sites (channels). Three of them are from the focal areas and the other three from nonfocal areas. Following [22] the signals were band-pass filtered between 0.5 and 65 Hz. Notch filter at 50 Hz was applied in the source itself. For each patient there are two to five ictal hours of recording (that is, one hour long six-channel signal contains one seizure along with pre and post-ictal recording. The entire data set is divided into one hour long segments. Except patient 2 for every patient there are also twenty four to twenty six hours of interictal recording. Further detail of the data can be seen in [16], [21].

#### B. Sparse and Dense Matrices

Step 1: Each of the three focal channel signals is operated upon by the first order difference operator with respect to time, which is nothing but $x(n) - x(n-1)$. Here $n$ is the time point.

Step 2: $\mathbf{M}$ is generated as in (1) by the difference operated focal channel digital signals, each 500 time points long. Obtain the following two matrices:

2(a) Generate the matrix $\mathbf{L}$ by convolving $\mathbf{M}$ with a $7 \times 7$ matrix with entries $1/49$ at all positions (moving average smoothing operation).

2(b) Generate the matrix $\mathbf{G}$ by first operating the Laplacian operator on $\mathbf{M}$, then taking the absolute value over the Laplacian operated $\mathbf{M}$ and finally smoothing out by convolving with a $7 \times 7$ matrix with all entries $1/49$.

Step 3: Perform elementwise multiplication $\mathbf{B} = \mathbf{L}.*\mathbf{G}$ (here we are following the MATLAB notation for elementwise matrix multiplication).

Step 4: A threshold is applied on the entries of $\mathbf{B}$ according to the following rule. Take 3 one hour long seizure free iEEG signals from the same patient and generate a class of $\mathbf{B}$s by sliding the 500 time point long signal window across one hour long signals with 50% overlap. Take the mean value of $\mathbf{B}$ in each case. Take the maximum of all the mean values over one hour long signals.

Step 5: Put 1 in the entry of $\mathbf{B}$ which is above the threshold determined in Step 4, and 0 otherwise. $\mathbf{B}$ has become a sparse matrix now. Opposite to sparseness is density. A matrix is x% dense if x% of its entries are nonzero. It is expressed in 0 to 1 scale as x/100.

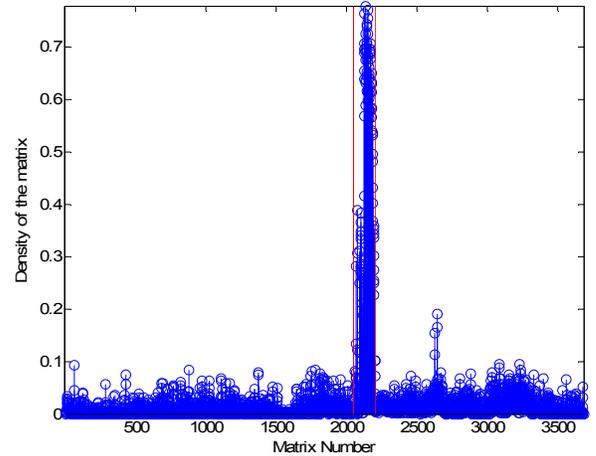

Fig 4. A $\mathbf{B}$ matrix (as in Step 5) of size 500 x 500 is slid with 50% overlap for focal iEEG signals of one hour duration. All the matrices are numbered successively during this one hour (abscissa). Density of the matrix goes very high during seizure (start and end indicated by vertical lines). A threshold on density will detect seizure.

Step 6: A patient specific threshold on $\mathbf{B}$ matrix density is decided by visually observing a couple of seizures (often 2 is a good number). In an unknown data if a $\mathbf{B}$ matrix crosses that threshold, the next $\mathbf{B}$ matrix is tested for having a density value, which is a certain fraction of the density of the preceding $\mathbf{B}$ matrix (which crossed the threshold first). If this following matrix has the density greater than or equal to the fractional density of the preceding then it is flagged as detection of a seizure. This fraction is again patient specific and determined by visual observation on a couple of seizures from the patient. The last point of the second matrix has been taken to be the seizure onset point.

Step 7: This step is for false detection avoidance only. When a seizure is detected in Step 6 we go back to Step 3 to calculate the mean and standard deviation of the matrix **B** in Step 3. Threshold is set for mean and standard deviation by taking slightly lower value for both than the values of **B,** which has exceeded the density threshold in Step 6. Subsequently, any **B** detected as belonging to seizure is checked if its mean and standard deviation in Step 3 has exceed the threshold. If not it is marked as an artifact and not a seizure.

### C. Detection Results

TABLE I
SUMMARY OF DETECTION BY MATRIX REPRESENTATION

| Patient number | Recorded iEEG in hours | # Seizure occurred | # Seizure detected | # False positives |
|---|---|---|---|---|
| 1 | 28 | 4 | 4 | 6 |
| 2 | 3 | 3 | 3 | NA |
| 3 | 29 | 5 | 4 | 1 |
| 4 | 29 | 5 | 4 | 0 |
| 5 | 29 | 5 | 5 | 0 |
| 6 | 27 | 3 | 3 | 4 |
| 7 | 28 | 3 | 2 | 0 |
| 8 | 26 | 2 | 2 | 2 |
| 9 | 29 | 5 | 5 | 5 |
| 10 | 29 | 5 | 0 | 0 |
| 11 | 28 | 4 | 3 | 0 |
| 12 | 29 | 4 | 4 | 4 |
| 13 | 26 | 2 | 1 | 0 |
| 14 | 28 | 4 | 4 | 8 |
| 15 | 28 | 4 | 3 | 0 |
| 16 | 29 | 5 | 3 | 0 |
| 17 | 29 | 5 | 5 | 1 |
| 18 | 30 | 5 | 3 | 3 |
| 19 | 28 | 4 | 4 | 28 |
| 20 | 31 | 5 | 5 | 0 |
| 21 | 29 | 5 | 5 | 0 |
|  | 572 | 87 | 72 | 62 |

Now in this subsection we will report the detection results by the novel matrix representation method described in the previous subsection. The detection results have been summarized in Table I. Since it was done on a publicly available data set we have reported the patient number in the first column. In the second column we have reported the total hour of recording for each patient. In the third column the number of seizure (ictal) hours (that is, one occurrence of seizure in one hour of recording) have been reported.

Recorded iEEG in hours – # seizure occurred = duration of interictal (seizure free) hour recording. The number of seizures detected successfully by the new matrix representation based method has been reported in the fourth column. Number of false positives in 24 hour long interictal (seizure free) recording has been reported in the last column. Patient 2 has an empty interictal data folder. That is why interictal recording has been shown as 'not available (NA).' However, the same patient has some hours of interictal recording within the ictal directory, which have been utilized in the next subsection. According to Table I the average false positive rate is 62/572 = 0.1084/hour. Only in case of patients 1, 14 and 19 the false positive rate is unacceptably high. However the average false positive rate is quite satisfactory. It has been reported to be 2.8/h in [23] compared to our 0.1084/h. 72 out of 87 seizures have been detected, that is, the detection accuracy is almost 83%. We ran our detection algorithm on all the seizures available in the data set, some of which even a trained epileptologist found hard to identify as seizure from the signals alone.

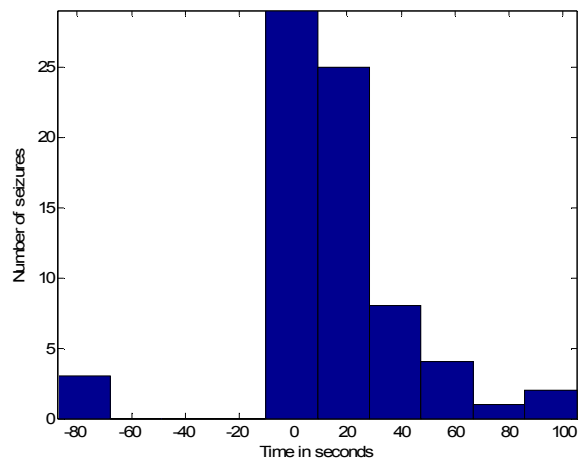

Fig 5. Histogram plot for detection latency of the seizures. The latency has been calculated from the epileptologist identified seizure onset time point.

Detection latency has been shown in Fig 5 in histogram form. The average detection delay across all detected seizures is 14.59 seconds, which can readily be improved by close to 3 seconds if we amend the last sentence of Step 6 in III(B) to "The first point of the first matrix ..." in place of "The last point of the second matrix …."

### D. Comparison

In this subsection we will compare our detection results in previous subsection with seizure detection by a standard machine learning algorithm [24]. Here a one class support vector machine (SVM) has been used to automatically detect epileptic seizures in iEEG signals. It is based on 'anomaly detection.' Here seizure and artifacts are treated as anomalies compared to artifact free signals. First, the SVM is trained on a seizure and artifact free signal. Then it is run on signals with seizure and artifacts. We have observed and utilized the fact that, seizure anomalies are of longer duration than the artifact anomalies, which has been the key to successful detection by this method in our implementation instead of making Bayesian inference as in [24].

The following two features have been extracted from the focal channel signals, after subjecting them to band-pass filtering between 0.5 and 65 Hz.
(i) Mean curve length (CL)

$$CL(n) = \log\left(\frac{1}{N}\sum_{m=n-N+2}^{n}|x(m)-x(m-1)|\right), \quad (3)$$

where $N$ is the number of time points in the window to slide. Here $N = 256$ and sliding is with 50% overlap. The window is slid within a larger window consisting of 921 time points, because we have taken one hour long signal segment consisting of 3600 x 256 = 921600 time points divided into 1000 bins, each 921 time points long (921600/1000 ≈ 921). The somewhat weird number 921 is constrained by the Python implementation.
(ii) Mean energy

$$E(n) = \log\left(\frac{1}{N}\sum_{m=n-N+1}^{n}x(m)^2\right). \quad (4)$$

A third feature called Teager energy was also utilized in [24], but it is quite similar to the mean energy and we checked that it was not improving the performance significantly. We went ahead with the above two features only.

One class SVM has been implemented by a Taiwan National University group and is available online freely [25] (we used the Python version). The same software was used in [24] for the one class SVM implementation. At the training stage one hour long seizure free signal was used, which was divided into 921 point long 1000 nonoverlapping segments. In each 921 point long segment 256 time point long window was slid with 50% overlap. In each window the features (i) and (ii) were calculated and entered into a two dimensional feature space.

256 time point window can be slid with 50% overlap in a 921 point long segment for 6 times. In each instance there will be a two dimensional feature vector and therefore there will be a total of 6000 points from a one hour long seizure free (baseline) signal. This is the training data set. During training the one class SVM classifies all these 6000 points into one class. During testing whatever feature vector falls on the other side of the SVM classifier is flagged as anomaly. An anomaly thus identified can be a seizure or an artifact. If in a one hour long signal the number of successive anomalies are high enough then it is called a seizure as shown in the anomaly histogram plot in Fig 6. In each seizure all three focal channels have been tested for occurrence of the seizure, but only the best result has been reported (seizure may not always occur in all the focal channels).

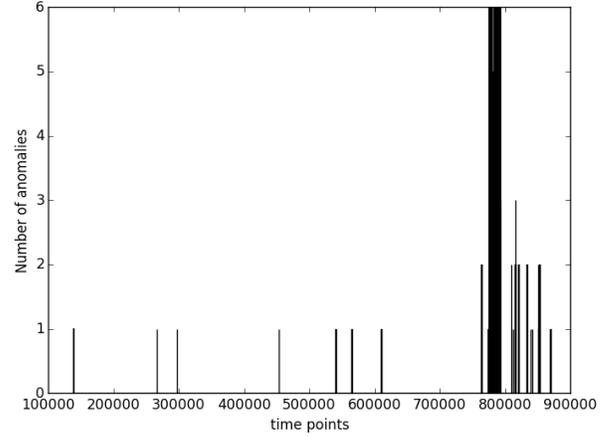

Fig 6. Anomaly histogram plot of ictal one hour focal channel recording from a patient with epilepsy. The bunch of closely spaced thick tall histograms indicates a seizure and its duration. The other smaller histograms are related to artifacts.

TABLE II
DETECTION SUMMARY BY THE ONE CLASS SVM

| Patient | # seizures | Detected | False positive | Mean latency | AuC |
|---------|-----------|----------|----------------|--------------|------|
| 1 | 4 | 4 | 1/h | 11.08 | 0.94 |
| 2 | 3 | 3 | 0.5/h | 11.72 | 0.97 |
| 3 | 5 | 4 | 0.5/h | 15.12 | 0.91 |
| 4 | 5 | 2 | 1.5/h | 18.43 | 0.82 |
| 5 | 5 | 5 | 0.75/h | 5.76 | 0.95 |
| 6 | 3 | 3 | 2.5/h | 12.02 | 0.77 |
| 7 | 3 | 1 | 0/h | 21 | 0.52 |
| 8 | 2 | 1 | 2/h | 10.2 | 0.7 |
| 9 | 5 | 5 | 2/h | 13.3 | 0.85 |
| 10 | 5 | 1 | 1/h | 10.1 | 0.4 |
| 11 | 4 | 3 | 1.75/h | 2.23 | 0.74 |
| 12 | 4 | 4 | 1/h | 12.76 | 0.87 |
| 13 | 2 | 1 | 1.5/h | 13.5 | 0.675 |
| 14 | 4 | 3 | 1.25/h | 13.2 | 0.71 |
| 15 | 4 | 3 | 2/h | 12.8 | 0.64 |
| 16 | 5 | 5 | 2.25/h | 3.56 | 0.88 |
| 17 | 5 | 4 | 4/h | 17.6 | 0.62 |
| 18 | 5 | 3 | 1.25/h | 10.1 | 0.75 |
| 19 | 4 | 4 | 1.25/h | 7.42 | 0.91 |
| 20 | 5 | 5 | 0.25/h | 8.67 | 0.97 |
| 21 | 5 | 5 | 1/h | 14.2 | 0.94 |
|  | 87 | 69 | 1.39/h | 11.66 |  |

False positive is for per hour. Mean latency is calculated for each patient. AuC = Area under (the ROC) curve. For patient 2 few interictal hours of recording were available within the ictal hour directory and false positives were calculated on them only.

Seizure detection results by the one class SVM has been summarized in Table II. 69 seizures out of a total of 87 have been detected compared to 72 detections by the matrix method

(Table I). False positives are pretty high by one class SVM at 1.39 seizures per hour (compared to 1.56/h in [24]) , whereas by matrix method the mean false positive rate across all the 87 seizures is 0.1084/h. Mean detection latency by one class SVM is 11.66s across all the seizures, whereas it is 14.59s for the matrix method. This can be improved to about 11.6s as explained at the end of III(C). However the detection latency by the matrix method is quite erratic as can be seen in Fig 5 and Fig 7. The detection latency by the one class SVM method is much more stable (Fig 7). However, the matrix method gives shorter detection latency compared to one class SVM for some patients. It even detects (predicts) seizures with negative latency for two patients.

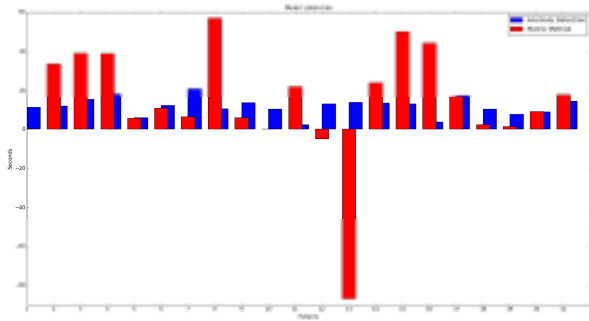

Fig 7. Bar plot of patient-wise average detection latency. Red is for matrix method and blue is for anomaly detection by one class SVM. The leftmost bar plot is for patient 1 and the rightmost is for patient 21. Abscissa is patient number and ordinate is detection latency in seconds.

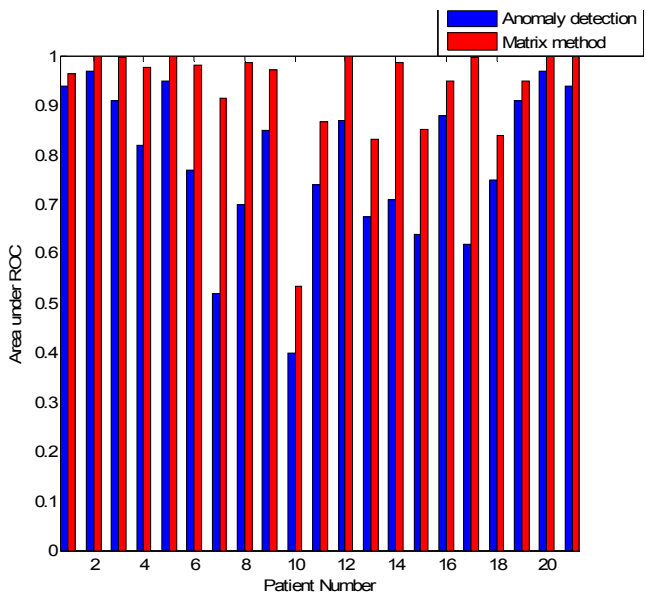

Fig 8. Histogram plots of area under the ROC curve for anomaly detection by one class SVM method (in blue) and by the matrix method for all the 21 patients. For the matrix method ROC curves were generated on density threshold alone.

Receiver operating characteristic (ROC) curve analysis is probably the most standard way to validate binary classification problems [26], like seizure detection. Area under the (ROC) curve (AuC) gives a performance measure of machine learning algorithms in case of binary classification [27]. Patient specific AuC by varying the threshold on the number of anomaly histogram height for seizure detection by one class SVM has been shown in the last column of Table II. In Fig 8 histogram plot of area under the ROC curve (AuC) have been shown for all the 21 patients. Red histograms represent the AuC for seizure detection by the matrix method and blue histograms show the AuC for seizure detection by the anomaly detection by one class SVM. There was no interictal data directory for patient 2 in the Freiburg data set, but there are interictal recording of few hours for this patient within the ictal directory, which we used for the false positive detection test. The ROC curves for the matrix method were generated on density threshold of **B** matrix alone (first part of Step 6 in III(B)). The threshold for making the matrix sparse was well defined and fixed (Step 4 in III(B)). The threshold for the fraction of density (opposite to sparseness) of the **B** matrix is fixed for any particular patient and was not varied for the ROC curve generation (last part of Step 6 in III(B)). It is clear from Fig 8 that the matrix method performed much better.

## IV. CONCLUSION

Linear Algebra is indispensible in signal processing. Matrix representation of digital signals readily brings Linear Algebra into the signal processing. Here we have shown how matrix representation can create a two dimensional image out of one or more number of digital signals. This image can be treated either algebraically or geometrically to extract meaningful information about the signals. In this work we have applied this new matrix representation of signals on iEEG signals of patients with epilepsy. We have shown how efficiently epileptic seizures can be detected from the geometric analysis of the matrix representation of multichannel focal iEEG signals. In future we plan to explore how automatic image threshold selection methods, like Otsu's algorithm [28] can improve the detection performance or can at least make it more objective. We have already used the Laplacian operator on this image in the current work. Other image segmentation techniques [29] can also be applied to extract information from this matrix.

One promising future research will be to apply a suitable wavelet decomposition on the matrix image [30] and encode the neural characteristics manifested in the associated signals in terms of those wavelet coefficients. For an appropriate window size this approach will enable one class of two dimensional wavelets to encode the neural correlates in the signals rather than one dimensional wavelets being applied on each single signal and then invoking additional techniques to interpret the ensemble findings. After the wavelet transformation machine learning algorithms are the natural candidates for classification according to the neural correlates of the signals.

**M** is a symmetric random matrix whose most of the entries will have small values and therefore can easily be made sparse by thresholding. In this paper we essentially presented a sparse random matrix encoding of iEEG signals, in which sparseness and randomness both go down during the seizure. Now, a big question will be, "Does it hold for some cognitive tasks as well?" If randomness and sparseness can be measured during cognitive tasks and they maintain a statistical relationship with the cognition then at least some of the brain functions can be coded by the randomness and sparseness of $\mathbf{M}$. Once such a representation of neural activities is possible the matrix $\mathbf{M}$ can be decomposed into basis matrices, whose neural correlate may offer us deeper insights into higher order cognitive processes. For example, we can have a clue to why a small stimulus, like a picture of a face, makes us recall instantaneously a whole of events and associations. Recent developments Compressed Sensing [31], [32] may give us efficient computational algorithms for the purpose.

We have already observed that first and second order differentiation with respect to time helps to enhance the contrast between the seizure part and the baseline part in iEEG signals [16], [17]. In Step 1 of III(B) we employed the first order differentiation for better detection. We have observed that other judicious combinations of first and second order differentiation prior to the matrix representation are capable of giving good results in seizure detection. This will be one of the future directions of our work.

In this work we have already observed that the highest eigenvalue of $\mathbf{M}$ goes up during a seizure in some patients. This needs further exploration in a patient specific manner. What physiological and pathological conditions are correlated with this phenomenon? It would be worth investigating if there is any correlation between ensemble phase synchronization [33] and highest eigenvalue of $\mathbf{M}$ in the focal channels before, during and after seizure.

ACKNOWLEDGMENT

This work was partially supported by a Department of Biotechnology, Government of India, Grant No. BT/PR7666/MED/30/936/2015. The authors like to thank Anupam Mitra for some critical suggestions, which have been incorporated into the paper.